\documentclass[10pt]{article}

\usepackage[margin=0.9in]{geometry}
\usepackage[T1]{fontenc}
\usepackage{times}
\usepackage{amsmath,amssymb}
\usepackage{booktabs}
\usepackage{graphicx}
\usepackage{float}
\usepackage{subcaption}
\usepackage{microtype}
\usepackage[colorlinks=true,linkcolor=blue,citecolor=blue,urlcolor=blue]{hyperref}
\usepackage{enumitem}
\setlist[itemize]{leftmargin=1.1em,itemsep=0pt,topsep=1pt}
\title{\bfseries GeoAgent: Contract-Constrained LLM Agents\\ for Seismic Data Processing}

\author{
  Zhilong Fang$^{1,*}$, Qi Zhang$^{1}$, Jujie Yan$^{1}$, Zhong Wang$^{1}$, \\
  Xinbin Gu$^{1}$, Ziyang Zhang$^{1}$, Hua Wang$^{1,*}$ \\[6pt]
  \small $^{1}$School of Resources and Environment, University of Electronic Science and Technology of China \\[2pt]
  \small $^{*}$ Corresponding author: Zhilong Fang, \texttt{fangzl@uestc.edu.cn}, Hua Wang, \texttt{huawang@uestc.edu.cn} \\[2pt]
  \small Live system: \url{http://www.geoagents.net/}
}

\date{October 2026}

\hypersetup{
  pdftitle={GeoAgent: Contract-Constrained LLM Agents for Seismic Data Processing},
  pdfauthor={Zhilong Fang, Qi Zhang, Jujie Yan, Zhong Wang, Xinbin Gu, Ziyang Zhang, Hua Wang},
  pdfsubject={LLM agents; seismic data processing; scientific workflow automation},
  pdfkeywords={LLM agents, seismic processing, tool use, workflow orchestration, reproducibility}
}

\begin{document}
\maketitle

\begin{abstract}
\noindent
Large language models (LLMs) can drive scientific software through function calling, but free-form tool use is fragile on long, data-dependent processing chains. Models omit arguments, invent units, feed the previous artifact into the wrong input, and report completion without a verifiable product. We present GeoAgent, an agent system for seismic data processing built on Madagascar's roughly 1{,}965 command-line programs. GeoAgent exposes 54 operator agents (51 generated from a declarative specification), retrieves from an index of 1{,}937 operators, and encodes the processing flow as a nine-stage state graph. Four explicit contracts---named artifacts, input resolution, parameter takeover, and artifact validation---move everything the graph can determine out of the model's discretion. A deterministic geometry tool parses the self-describing SPS/RPS/XPS and SEG-Y headers, cross-checks eight consistency conditions, and reproduces the reference shell recipe exactly. The system runs in containers with session- and task-level isolation, bounded execution, and defenses against silent failures. On a 2-D land dataset, the nine-stage pipeline completes in about six to eleven minutes and yields a verified artifact at every stage, while retrieval reduces per-turn tool context from about 12k to 2k tokens. We report the design, deployment experience, and motivating failure modes.
\end{abstract}

\section{Introduction}

Seismic processing turns raw field records into subsurface images through long chains of steps: loading, geometry, muting, filtering, velocity analysis, normal-moveout (NMO) correction, stacking, and migration. Madagascar~\cite{fomel2013madagascar} makes this capability open: roughly 1{,}965 \texttt{sf*} programs, an RSF format pairing a text header with a binary hypercube, and reproducible SCons recipes. The price is expertise---which operator to use, how operators compose through pipes, which parameters are mandatory, and how data flows between stages.

Function calling in LLMs offers a natural interface to such software. The common approach wraps every operator as a tool, injects all descriptions into the context, and lets the model choose calls and parameters~\cite{schick2023toolformer,yao2023react,qin2024toolllm}. On seismic flows this fails characteristically. Descriptions crowd out the task; the model omits required parameters or invents units; it assumes the previous output is the next input, although NMO needs common-midpoint (CMP) gathers from velocity analysis while the velocity field enters as an argument; it declares completion when the product is missing, mistyped, or empty; it loops without bound; and concurrent users interfere through global state.

Retrieval-augmented generation (RAG)~\cite{lewis2020rag} reduces context pressure by retrieving a few operators per turn, but these are candidates, not an executable dependency graph. Hard-coded workflows have the opposite weakness: deterministic but inflexible.

GeoAgent takes a middle position: \emph{let the model decide locally, and let the graph decide globally}. The LLM still chooses tools and reasons about data within a stage. Stage order, inter-stage inputs, mandatory parameters, and acceptance criteria are specified by an immutable stage graph and enforced by the runtime. Our contributions are:
\begin{itemize}
  \item a five-layer, containerized architecture serving standard, RAG, and pipeline modes over one execution layer;
  \item a five-module operator-agent template with spec-driven generation (54 agents);
  \item four executable contracts on a nine-stage graph that replace the ``previous output is the next input'' assumption;
  \item a deterministic, self-description-driven geometry builder for SPS/RPS/XPS and SEG-Y data, validated against the reference recipe;
  \item an account of deployment, isolation, silent-failure defenses, and the failures that motivated each mechanism.
\end{itemize}
This is a system and experience report, not a controlled benchmark: measurements come from one 2-D land dataset and development iterations. A live deployment is available at \url{http://47.109.133.239/}.

\section{Related Work}

Tool-using LLMs interleave reasoning with external calls~\cite{yao2023react,schick2023toolformer}; scaling to thousands of APIs has been studied through API instruction tuning~\cite{qin2024toolllm}, documentation retrieval~\cite{patil2023gorilla}, and model orchestration~\cite{shen2023hugginggpt}. RAG~\cite{lewis2020rag} supplies context but not execution guarantees. Scientific agents exist for chemistry~\cite{bran2024chemcrow,boiko2023coscientist}, and geoscience has domain LLMs for knowledge tasks~\cite{deng2024k2,zhou2024geogalactica}; we instead orchestrate an existing software package. Workflow engines such as LangGraph~\cite{chase2024langgraph} and SCons recipes provide stateful, reproducible graphs. GeoAgent adds contracts binding model behavior to graph facts: which artifact feeds which input, which arguments are fixed, and what counts as completion. Multi-agent frameworks~\cite{wu2023autogen} focus on conversation; we constrain one model's local autonomy.

\section{System Overview}

\begin{figure}[t]
  \centering
  \includegraphics[width=0.80\linewidth]{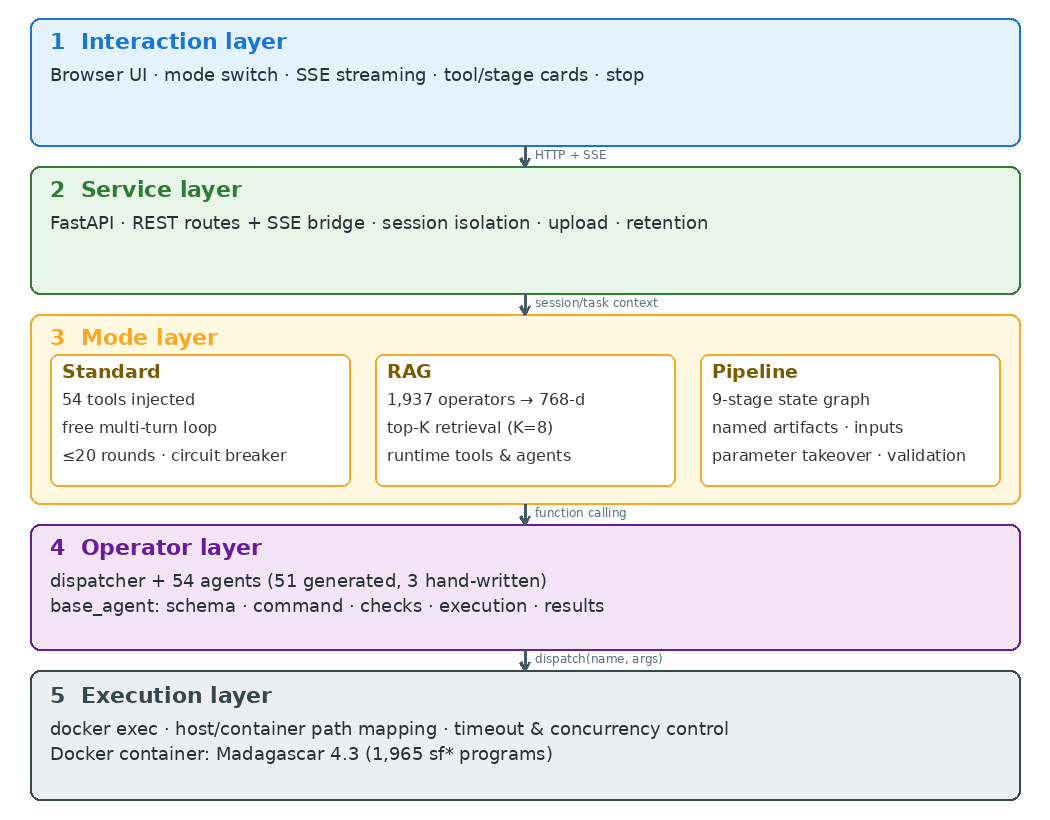}
  \caption{The five-layer architecture. All modes share the operator and execution layers. Standard mode injects 54 tools; RAG mode retrieves from 1{,}937 indexed operators; pipeline mode runs a nine-stage state graph with artifact, input, parameter, and validation contracts.}
  \label{fig:arch}
\end{figure}

GeoAgent has five layers (Figure~\ref{fig:arch}). The \emph{interaction layer} is a browser UI that switches modes and renders server-sent events: model text, tool cards, results, inline images, and stage summaries. The \emph{service layer} is FastAPI with REST routes and an SSE bridge: a producer thread runs the model and operators, writes events to a queue, and an asynchronous response drains it. A stop request sets a flag without closing the stream. The \emph{mode layer} holds three runtime-switchable schedulers. The \emph{operator layer} is a registry and agent pool reached through a dispatcher. The \emph{execution layer} runs commands in a Docker container hosting Madagascar~4.3.

Deployment uses three containers: a reverse proxy, the application, and the operator container. The proxy is the only external entry point; the operator container publishes no port. The host data directory is mounted at the same path in both containers, so path conversion cannot drift. Our deployment has four virtual CPUs and about 7.3~GiB of memory.

\section{Operator Agents}

Each Madagascar program is wrapped by an agent built from a common base class with five modules: (1) function metadata exporting a name, description, and JSON Schema; (2) command assembly mapping structured arguments to the native shell command, appending optional arguments only when supplied; (3) argument validation for required fields, file existence, and output-path normalization; (4) sandboxed execution with a timeout; and (5) structured result packaging with status, message, output, RSF header, and host-visible paths.

Agents are generated from a declarative specification that declares the program, tool name, description, schema, command template, file parameters, extra arguments, output extension, and timeout. A generator emits the agent class, schema, and registry entry, so changing an operator means editing a specification. The pool contains 54 tools: 51 generated agents and three hand-written tools---geometry construction, survey-data probing, and a whitelisted shell sandbox that accepts only \texttt{sf*}, \texttt{awk}, and similar per-segment constraints.

Execution goes through \texttt{docker exec}. Host and container paths are converted by prefix mapping, and every path returned to the model is a host path, so the model never reasons about two namespaces. Results include RSF dimensions and sampling when available, letting later stages check dimensional expectations.

\section{Three Interaction Modes}

\paragraph{Standard mode.} All 54 tools are injected and the model runs a free multi-turn loop, up to 20 rounds. Results stay addressable, so a user can say ``band-pass the previous result.'' If the same tool fails three times in a row, a circuit breaker forces a strategy change or a stop. The mode is flexible but has no global data-flow guarantee.

\paragraph{RAG mode.} An offline index parses operator manual pages into metadata encoded as 768-dimensional vectors; it covers 1{,}937 operators. At run time the request is encoded, compared by cosine similarity, and the top $K$ operators (default $K{=}8$) are instantiated as agent classes on the fly; their schemas are the only tools the model sees. Per-turn tool context falls from roughly 12k to 2k tokens, allowing smaller models to participate. Retrieval still returns candidates, not a contract.

\paragraph{Pipeline mode.} The flow is a directed state graph implemented with LangGraph~\cite{chase2024langgraph}. A stage entry node resolves inputs, builds a stage-specific prompt, and calls the model with a stage-specific tool set. A tool node executes calls and records successes, failures, and candidate products. A validation node applies the acceptance contract, and a summary node advances. Each stage allows eight tool rounds and three attempts. The model sees only the current transcript plus one-line summaries of completed stages; on success the transcript is cleared, so context does not grow with flow length.

\section{Contracts on the Stage Graph}
\label{sec:contracts}

\begin{figure}[t]
  \centering
  \includegraphics[width=0.86\linewidth]{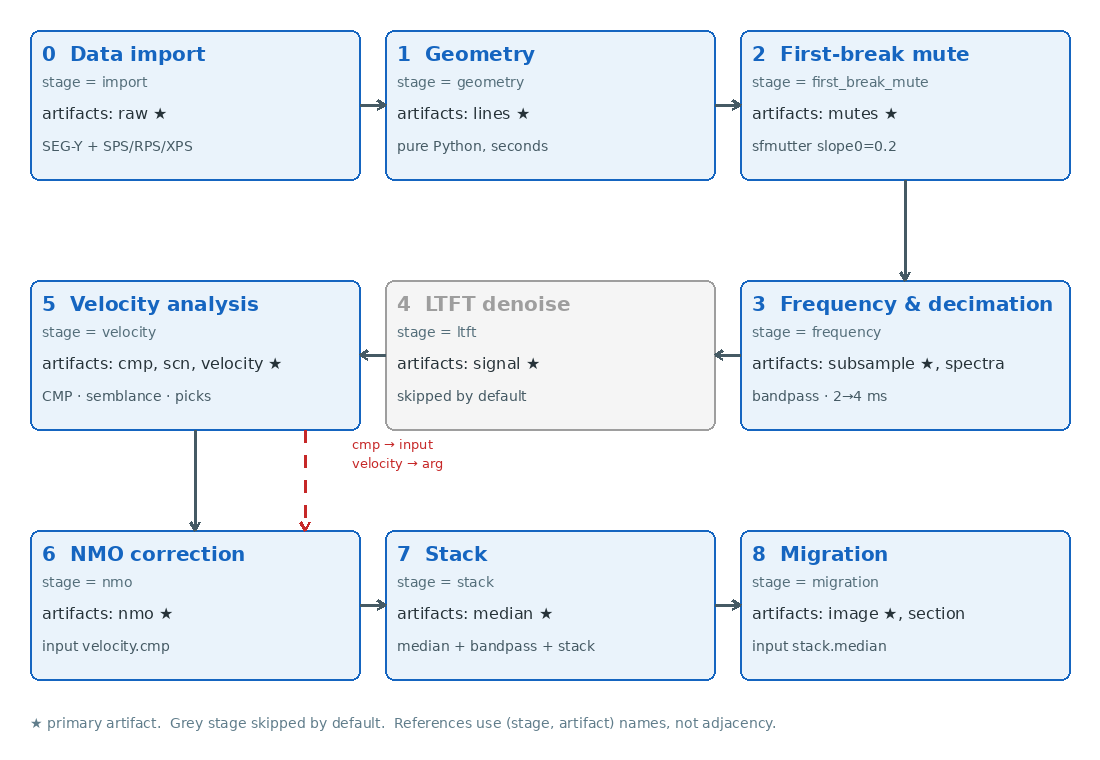}
  \caption{The nine-stage pipeline. Each stage declares named artifacts and an explicit input source. NMO consumes the CMP gather from velocity analysis and receives the velocity field as an injected argument (red dashed arrows); neither relation is adjacency-based. LTFT denoising is defined but skipped by default, and the skip is reflected in the velocity stage's declared input.}
  \label{fig:pipeline}
\end{figure}

A stage specification is immutable and declares its name, prompt, tools, default skip flag, input source, named artifacts, parameter policy, required tools, artifact type, and retry bound. Order follows declaration, but references use names, not indices, so stages can be inserted, removed, or reordered. Four contracts make the graph authoritative (Figure~\ref{fig:pipeline}).

\paragraph{Named artifacts.} A stage may declare several products, each with a key and a producing tool, and may mark one as primary. Downstream stages reference a product by (stage, key). Velocity analysis, for example, declares CMP gathers \texttt{cmp}, a semblance panel \texttt{scn}, and a velocity field \texttt{velocity}; NMO uses \texttt{velocity.cmp} as data and \texttt{velocity.velocity} as a parameter. A product may carry a deterministic post-step: the CMP gather from \texttt{sfshot2cmp} has its offset-axis header corrected by \texttt{sfput} before registration, matching the reference recipe.

\paragraph{Input resolution.} Every stage declares \texttt{input\_source} as a (stage, key) pair; the entry node resolves the path from the artifact registry. Only an undeclared stage falls back to the nearest completed stage's primary product. This expresses non-linear dependencies directly.

The graph is checked when it is built: a missing stage, undeclared key, or reference to a default-skipped stage aborts construction instead of failing mid-run.

\paragraph{Parameter takeover.} Parameters have three ownership levels. \emph{Fixed} values are stage-set and cannot be changed by any source. \emph{Guided} values have defaults, bounds, or enumerations and may be model-locked: the model cannot change them, but a user can at an approval gate. \emph{Free} values pass through; a closed policy forbids undeclared arguments. Two declarations bind data to arguments: \texttt{inject\_args} fills an argument from an upstream artifact, and \texttt{bind\_input\_tools} forces a tool's input to the stage input. Resolution order is model values, guided defaults, user intent, explicit user patch, fixed values, and finally injection. Because the last two run last, fixed values and data-flow bindings cannot be overridden by accident---a property of order, not of prompt.

\paragraph{Artifact validation.} A stage passes when
\begin{equation}
\mathrm{valid}(P) = \mathrm{exists}(P) \;\wedge\; \bigl(\forall t \in R:\; N(t) \ge 1\bigr) \;\wedge\; \mathrm{ext}(P) \in K,
\label{eq:valid}
\end{equation}
where $P$ is the product path, $R$ the required tools, $N(t)$ the successful executions of tool $t$, and $K$ the allowed types. QC images and undeclared outputs cannot masquerade as the product. On failure, the summary and guidance enter the next attempt; when retries are exhausted the pipeline stops with a structured error instead of continuing on bad data.

\section{Self-Describing Survey Data and Deterministic Geometry}

Survey files (SPS/RPS/XPS) are fixed-width text, and column positions and header lengths vary. Hard-coding ``column 8 and 9'' or ``skip 20 lines'' produces uniformly shifted coordinates while array dimensions still look plausible---a silent error. GeoAgent reads each file's field table instead. Header lines beginning with \texttt{H} describe fields and column ranges, e.g., \emph{Map grid easting 47--55}; the parser builds a field-name-to-column map and finds the first non-header record. Coordinates are sliced by field name, and a fallback to whitespace splitting is reported if the table is absent.

It also parses key--value acquisition parameters and the 3{,}200-byte textual and 400-byte binary SEG-Y headers, then runs eight cross-checks: station ranges against coordinate counts, XPS records against SPS points, regularity of traces per shot and receiver step, file-length divisibility, sampling plausibility, sweep-band presence, and measured versus nominal bin spacing. A failed check is reported rather than used to build a wrong geometry.

Construction is pure Python inside the container. XPS fixes each shot's receiver station range; station numbers map to RPS indices; source coordinates are read per shot; meters become kilometers; and matrices are assembled with receivers varying fastest. Offsets follow
\begin{equation}
\mathrm{offset}(i,j) = \sqrt{\bigl(SX_{ij}-RX_{ij}\bigr)^2 + \bigl(SY_{ij}-RY_{ij}\bigr)^2},
\label{eq:offset}
\end{equation}
and the result is written as an RSF cube $[n_{\text{trace}}, n_{\text{shot}}, 1]$. The data volume is reordered into trace-by-shot-by-time, removing per-shot auxiliary traces to align with the geometry. On the test dataset all eight checks pass: 251 shots, 782 receivers, 25~m and 50~m spacings, 2~ms sampling, and an 8--95~Hz sweep. The output agrees element-wise with the original shell recipe (maximum absolute difference 0), in seconds.

\section{Reliability and Isolation}

\paragraph{Two-level isolation.} Each browser session owns a session object holding its gateways, history, stop event, and round token. A per-session lock allows one round at a time, and a monotonic token prevents a superseded producer from writing into a newer round. A context object carries session, task, and directory, propagated through a context variable. Products go to \texttt{sessions/<sid>/outputs/<task>}; retention removes a whole session tree.

\paragraph{Bounded execution.} A global semaphore limits concurrent operators; auxiliary commands are exempt to avoid self-deadlock, and per-operator threads are capped. Timeouts use container-side GNU \texttt{timeout} on the whole process group, so orphans do not survive. A stop request interrupts queued semaphore acquisition. A watchdog releases a wedged round's lock and detaches its gateway, so the stale producer writes into a discarded object.

\paragraph{Silent failures.} Several failures report success at the process level. A plot with a non-ASCII title exits with status zero while writing a 0-byte PNG; an axis longer than 32{,}767 pixels fails outright; a blank image is still a valid file. GeoAgent strips non-ASCII from plot text, clamps oversized axes, and treats an empty PNG as failure. Model-supplied output paths, relative or absolute, are reduced to a task-local filename, preventing products from escaping retention. Transient model-service errors are retried with bounded exponential backoff.

\section{Deployment Experience}

We ran the system on a 2-D land dataset (251 shots, 782 receivers) with a mid-sized GLM model behind an OpenAI-compatible API; the model is replaceable. The flow covers import, geometry, first-break mute, frequency analysis and decimation, velocity analysis, NMO, stack, and migration; LTFT denoising is defined but skipped by default.

End-to-end runs take about six to eleven minutes depending on model latency. Every stage returns a non-empty product path, with mutually consistent dimensions: geometry $1501\times282\times251$ (time, offset, shot), after decimation $751\times282\times251$, CMP gathers $751\times71\times1285$ (trace, offset, midpoint), NMO gathers of the same shape, and a stacked section of $751\times1285$. Geometry matches the reference exactly, and RAG reduces per-turn tool context from about 12k to 2k tokens. Table~\ref{tab:failures} lists observed failures and the structural mechanism that prevents each.

\begin{table}[t]
\centering
\footnotesize
\caption{Failure modes observed during development and the mechanism that prevents recurrence.}
\label{tab:failures}
\begin{tabular}{@{}p{0.42\linewidth}p{0.52\linewidth}@{}}
\toprule
Observed failure & Structural remedy \\
\midrule
NMO receives the velocity field as data; output stays one gather and the stack collapses to one trace. & \texttt{input\_source=(velocity,cmp)}; the velocity field is injected into \texttt{velocity=}. \\
Model registers an arbitrary RSF as the velocity field. & The \texttt{velocity} artifact must be produced by the declared picking tool. \\
Model synthesizes a 1-D ramp as the migration pick. & \texttt{pick} is injected from \texttt{velocity.velocity}. \\
A tool is invoked twice (forward and inverse transform); input binding rewrites the second call. & Bind input only for tools invoked once per stage and reading the stage input each time. \\
Model writes an absolute path outside the session tree; plots silently yield 0-byte PNGs. & Paths are reduced to task-local filenames; plot text is scrubbed, axes clamped, and blank images treated as failures. \\
\bottomrule
\end{tabular}
\end{table}

\paragraph{What the contracts buy.} The failure table is the main empirical result. Each row was a reproducible defect, and each remedy is structural: it removes a degree of freedom rather than adding an instruction. Instructions were not sufficient---telling the model not to use the velocity field as input did not stop the behavior, whereas declaring the input source did. Freedom where the graph has no opinion---choosing a QC plot, when to summarize, a band adjustment---remains useful and is preserved.

\begin{figure}[t]
  \centering
  \begin{subfigure}{0.49\linewidth}
    \centering
    \includegraphics[width=\linewidth]{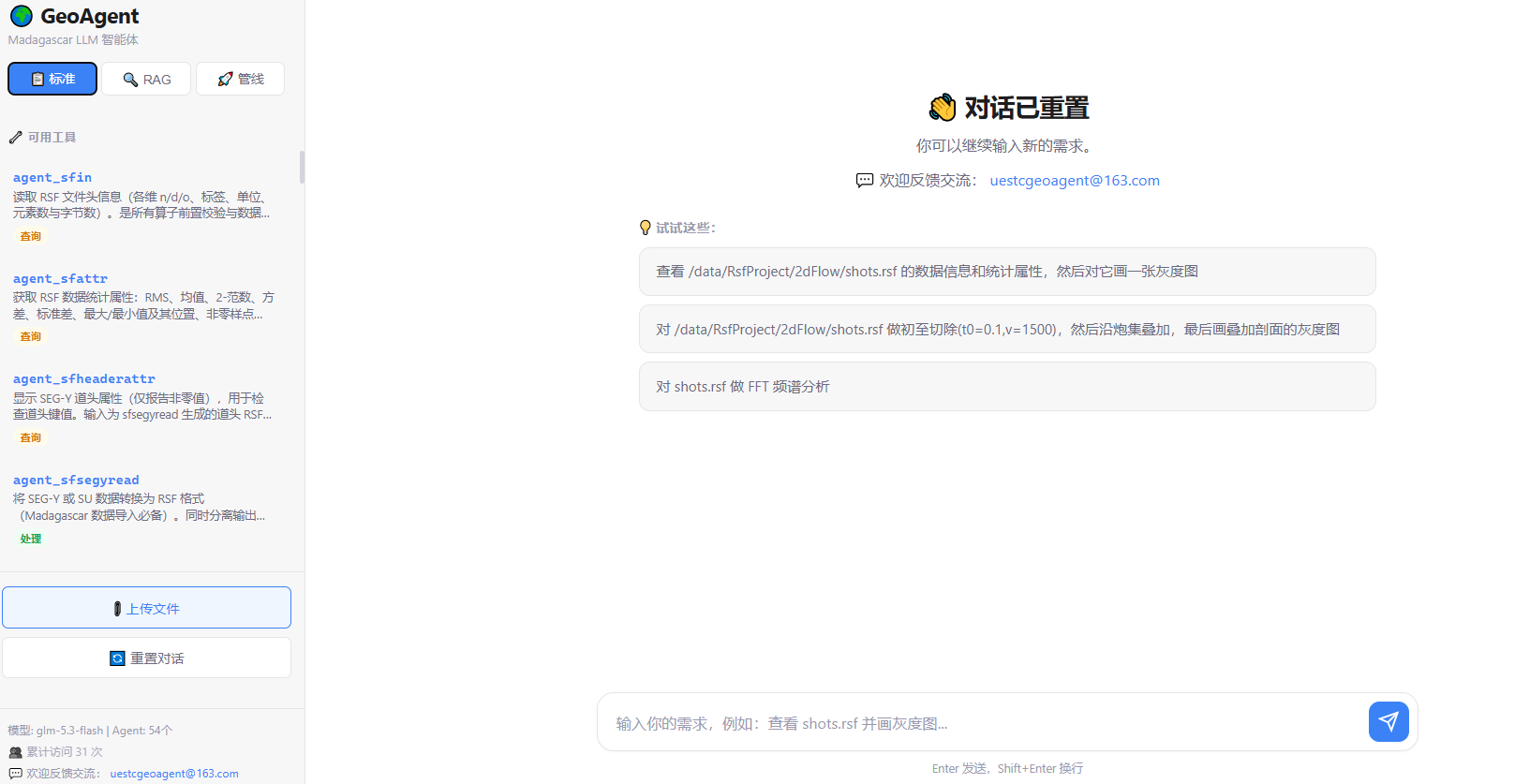}
    \caption{Standard mode}
  \end{subfigure}\hfill
  \begin{subfigure}{0.49\linewidth}
    \centering
    \includegraphics[width=\linewidth]{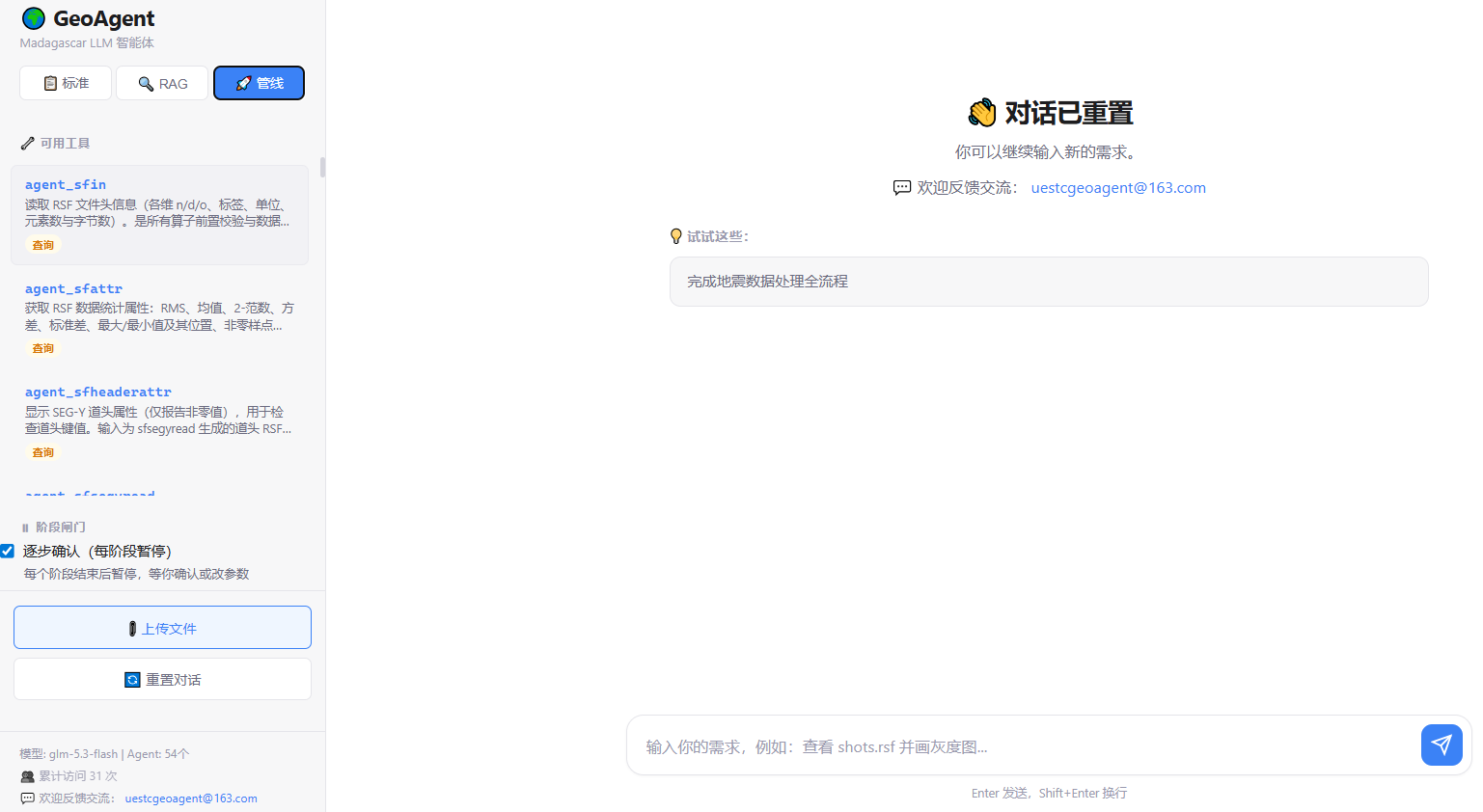}
    \caption{Pipeline mode with the per-stage confirmation gate enabled}
  \end{subfigure}
  \caption{Entry screens of the deployed system. The standard panel lists the injected tools and offers example requests; the pipeline panel starts the nine-stage graph and can pause after each stage for review, retry, skip, or parameter changes.}
  \label{fig:ui}
\end{figure}

Figure~\ref{fig:ui} shows the deployed interface. The standard sidebar exposes all 54 tools; the pipeline sidebar exposes only the current stage's tools and offers an optional per-stage confirmation gate. The interface is shown in its Chinese localization.

\section{Discussion and Limitations}

GeoAgent has no planning layer: input does not yet reorder or select stages, and the graph is fixed per dataset family. Migration is enabled, but its stability and image quality are still under validation; mute and LTFT parameters need calibration on more datasets. Some C programs are not built, so FK filtering and surface-consistent amplitude corrections are unavailable. Concurrency limits, memory, and checkpointed pause-and-resume are still being engineered; a user gate supports modifying, retrying, skipping, or jumping stages.

The evaluation is deliberately modest: one dataset, one model family, and engineering measurements rather than a controlled study. A proper benchmark would compare task success, operator coverage, parameter correctness, and human effort across models and datasets, with repeated runs and contract ablations. We present the architecture and failure taxonomy as the contribution.

\section{Conclusion}

Free-form LLM tool use and fixed scientific workflows fail in complementary ways. GeoAgent separates the two responsibilities: the model reasons locally within a stage, while an immutable graph with named artifacts, explicit inputs, parameter ownership, and artifact validation decides what must be true. A deterministic, self-description-driven geometry builder removes a class of silent data errors, and container isolation and silent-failure defenses make the system usable by multiple sessions. The result is a practical natural-language interface to a large seismic package that is reproducible at the stage level and honest about failure.

\end{document}